# TESTING THE USE OF THE PRINCIPAL COMPONENT ANALYSIS METHOD TO DETECT SLIGHT CHANGES IN NITROGEN AND OXYGEN SPECTRA INDUCED BY A CHANGE IN DISCHARGE CONDITIONS

M. F. Yilmaz[1], A. Fkereen[2], Y.Danisman[3], J. Larour[4]

[1] Independent Researcher, Freemont, CA, USA
[2] Engineering Department, Yıldırım Beyazit University, Ankara, Turkey
[3] Department of Mathematics, University of Oklahoma, Norman, OK, USA
[4] Laboratoire de Physique des Plasmas (LPP), Ecole Polytechnique, UPMC, CNRS, Palaiseau, France

**Abstract:** In this work, the influence of crossed magnetic fields (B=100 Gauss) on the UV-vis and NIR spectra and breakdown voltages of nitrogen and oxygen at a pair of parallel plane copper electrodes with a spacing d=2 cm and diameter R= 2.2 cm are studied. Working pressures of the gases are kept between 0.1 and 1 Torr. The breakdown voltage measurements across the electrodes are conducted by Tektronix P6015 high voltage probe connected to a Tektronix 2430A oscilloscope. The Spectra of discharge plasmas in the absence and presence of magnetic fields are recorded between 200-1100 nm. by AvaSpec-ULS3648. In order to analyse observed spectra of nitrogen and oxygen plasmas, one of the pattern recognition techniques of principal component analysis has been employed. Results of principal component analysis shows that the presence of magnetic field cause the plasma particles to move in condensed way for the oxygen but not for the nitrogen. Principal component spectra shows that small amount of the cross magnetic field results in the Stokes linear polarization of the most of the oxygen and NIII (337 nm) of the Nitrogen species in the plasmas. The non-local thermal equilibrium based spectroscopic modeling of nitrogen and oxygen plasma gives that the electron temperature of nitrogen and oxygen are Te=25000 and 5000 Kelvin respectively.

I. **Introduction.**

Glow discharge plasmas are mainly generated by two electrodes which are inserted to a glass tube and connected to a power supply. At the point when a sufficient high potential distinction is connected between two electrodes set in a gas at low pressures of 0.1-1 Torr, will separate the gas into positive particles and electrons, offering ascend to a gas release. The electrons are accelerated towards the anode by the electric field to collide with the gas atoms. The ions which are accelerated by the electric field move towards the cathode and the collision of ions with the cathode results in secondary emission. Inelastic collisions of electrons leading to excitation and following de-excitations resulting in the emission of radiation are responsible for the characteristic name of the 'glow' discharge [1].

Plasmas consist of electrons, ions and neutral species. Glow discharge plasmas are considered to have a cold temperature, electron temperatures of weakly ionized non-thermal plasmas are relatively higher than those of ions ($T_{electron} \gg T_{ion} = T_g = 300 \ldots 10^3$ Kelvin) and electron densities are around $n_{electron} = 10^{16} cm^{-3}$. For that reason, the thermal motions of the ions and neutrals can be ignored, and the gas in cold plasmas is mainly heated by the thermal motions of the electrons due to inelastic collisions. Another characteristic of the non-thermal plasmas is that the velocity distributions of species do not follow the Maxwell-Boltzmann statistics [2].

Cold nitrogen and oxygen plasmas generated by the glow discharge method are used in bio-medical and industrial applications, waste water managements, decontamination and sterilization, material and surface treatment, carbon and nanotube growth etc. where heat is not desirable. Since air is mainly composed by nitrogen and oxygen, plasmas of these gases play very important role in understanding air pollution. UV–vis and NIR emission spectroscopy are non-invasive methods and, they can be used for diagnosing cold plasmas. It can enable to identify an active atomic or molecular species, and non-thermal modeling of these spectra can provide insight for the plasma atomic and molecular processes of excited atomic and molecular states [3, 4].

Principal component analysis (PCA) is one of the most common techniques used for dimension reduction and investigating hidden structures of multivariate dataset. PCA reduces the dimension of data by projecting it onto a space spanned by the vectors called principal components. These vectors are obtained successively and correspond to the maximum variance of the remaining data in each time. PCA has been applied in many areas, such as medicine, robotics and remote sensing. It has also many applications in spectroscopy, especially in unmixing species and decomposing overlapped spectral lines of UV-VIS-NIR spectroscopy to extract the spectral fingerprints. Besides, it is used in spectroscopy of astrophysical and laboratory plasmas to extract the plasma parameters and compositions of ion species and their interactions with electron beams [5].

In this study, the effects of magnetic field on the nitrogen and oxygen glow discharge plasma are analyzed by means of UV-vis and NIR spectroscopy. The experimental part is based on the oscilloscope measurements and UV-VIS spectroscopy. Pattern recognition technique of the principal component analysis (PCA) is applied to spectral dataset of nitrogen and oxygen plasma to extract the effects of magnetic fields [6]. The second chapter gives the details of experiments performed and the third chapter studies PCA. The fourth chapter discusses non thermal equilibrium (LTE) spectral modeling of the plasma. The fifth chapter is the discussion part and the conclusions are given in the final chapter.

II. Experimental setup.

The experiments described in this article are performed in D.C glow discharge plasma using the nitrogen and oxygen. Figure.1 shows the experimental setup. The breakdown voltage is formed in parallel plate configuration of copper electrodes with a diameter of 2.2 cm and spacing 2 cm, housed in a cylindrical glass tube vacuum chamber of 4.8 cm in diameter and 10 cm in length, with inlet gas discharge and vacuum outlet. The discharge chamber is applied using 100-2000 VDC power supply, and the current was 5.0 mA. During all measurements, a continuous flow of a gas through the discharge chamber was maintained. The nitrogen and oxygen gases are used as working gas and were fed to the chamber through controlled flow rate. The discharge chamber was pumped down by a mechanical vacuum pump to base pressure of 0.1-1 Torr. The spectral of breakdown plasmas are recorded by AvaSpec-ULS3648 between 200-1100 nm.

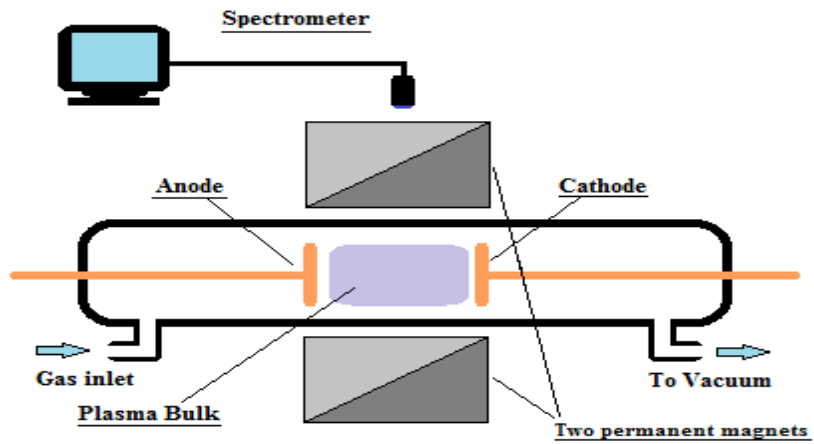

Fig 1 *Experimental setup*

Figure 2 shows the comparative chart for breakdown voltages of nitrogen and oxygen plasmas in the presence and absence of the magnetic fields of 100 Gauss. Magnetic fields are generated using two permanent magnets. The resulted data were on the right side of the Paschen curve of the oxygen. The received breakdown voltages are higher, which are expected due to distance of the electrodes. The influence of cross magnetic field causes a slight increase (2 %) of the the breakdown voltages [8].

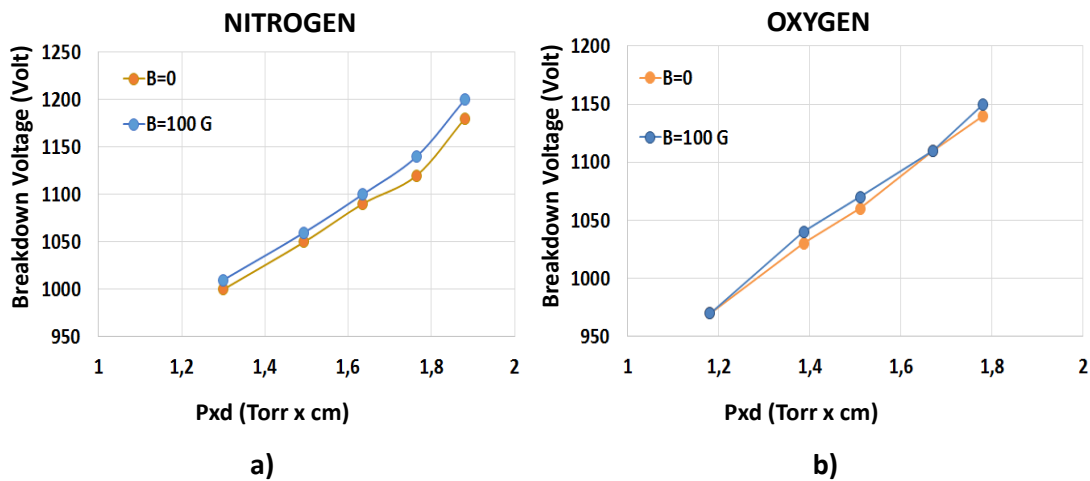

Fig 2 *Comparison of breakdown voltages of a) nitrogen and b) oxygen in the absence and presence of crossed magnetic field of 100 Gauss*

Figure 3 and 4 illustrates the time dependence of voltage and current measurements. The measurements show that the pulse timing of nitrogen and oxygen are decreased from 0.240 to 0.210 sn and 0.320 to 0.210 sn, respectively. So the magnetic field is more effective on pulse timing of Oxygen compared to the nitrogen [9].

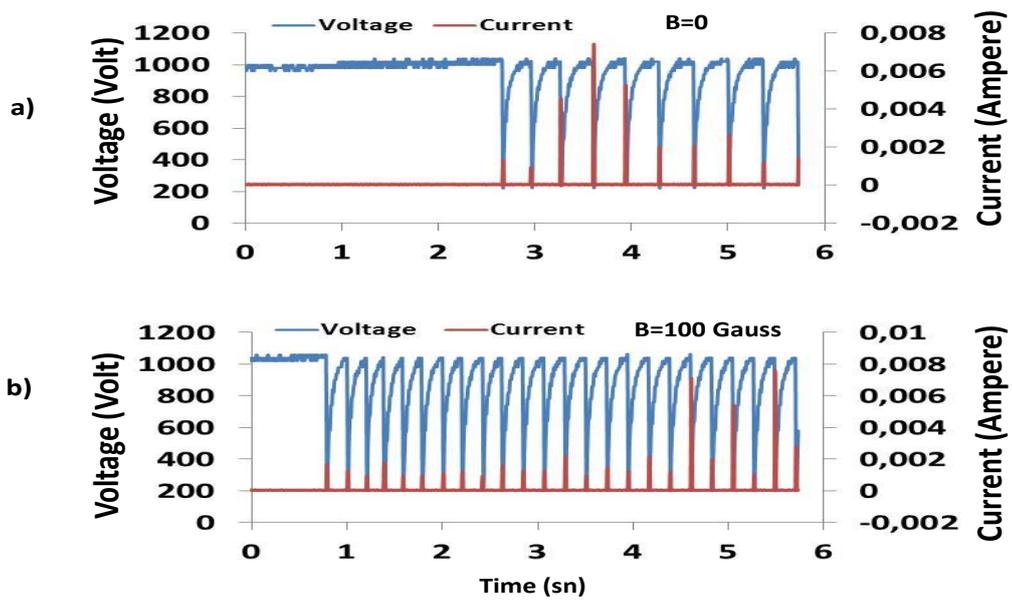

Fig.3 Oscilloscope data of Nitrogen a) B=0, b) B=100 Gauss at 1.3 Pxd (Torr x cm)

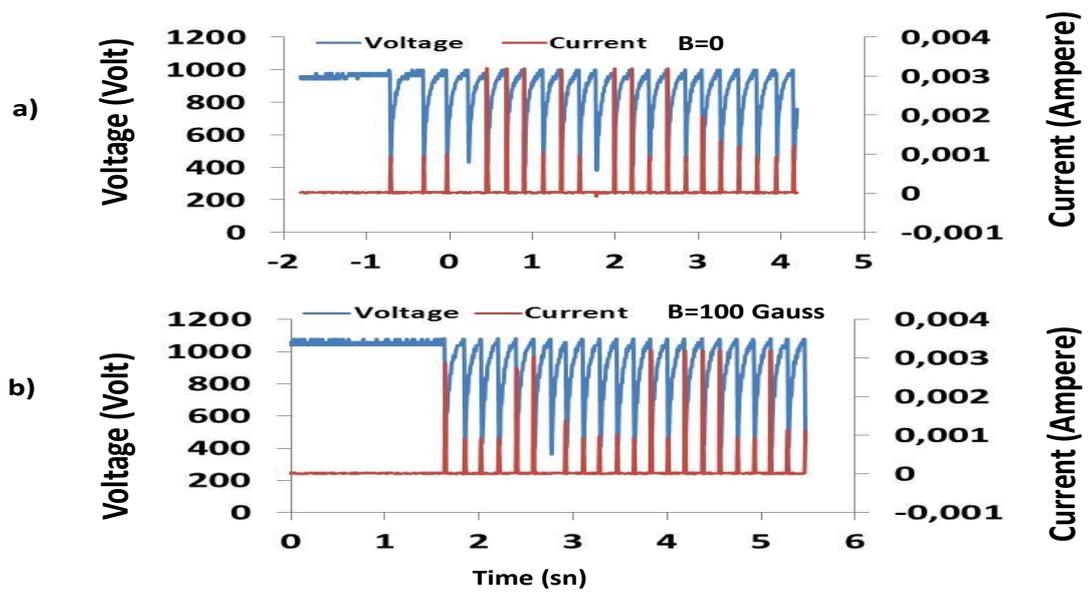

Fig.4 Oscilloscope data of Oxygen a) B=0, b) B=100 Gass at 1.18 Pxd (Torr x cm)

In figure 5 the images of nitrogen and oxygen plasmas are illustrated. Nitrogen plasma has a brighter and larger volume on the anode side. However, the volume of the glow of the cathode side and faraday dark space of the oxygen appears to be larger [1].

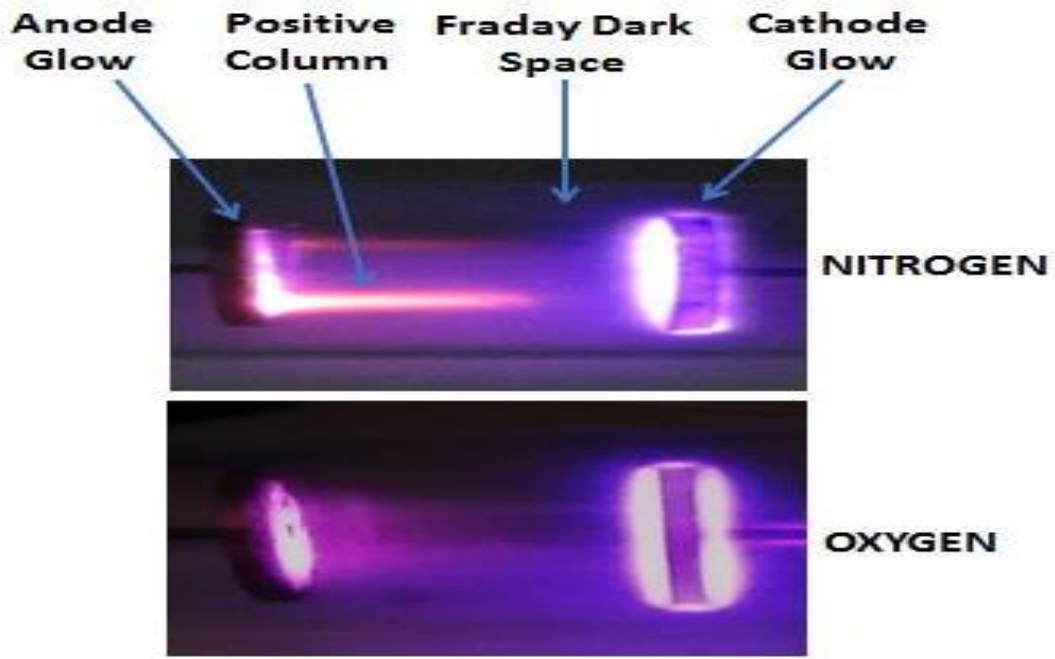

*Fig.5 Image and nitrogen and oxygen discharge plasma.*

The obtained spectra of discharge plasmas of nitrogen and oxygen are shown in figure 7. The spectra show oxygen is noisier than the nitrogen. Nitrogen spectra have a more intense radiation in the UV and Vis, however oxygen has it in Vis and NIR region for the considered experimental conditions. The most intense radiations of nitrogen and oxygen are 337 and 777 nm, respectively. Oxygen transitions at 845 nm are also observed [10, 11].

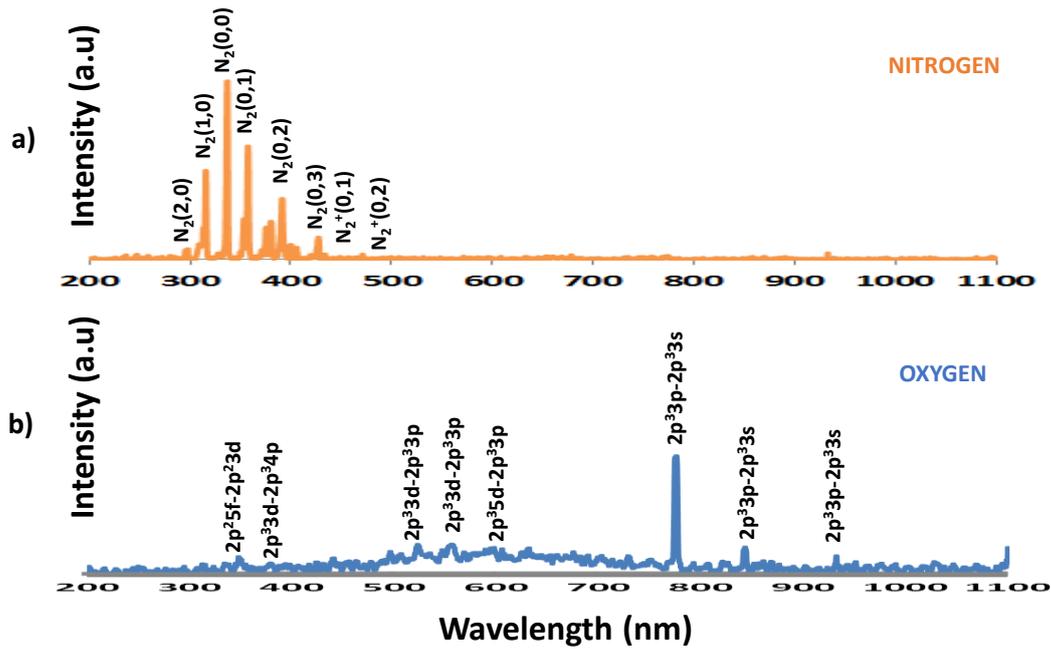

Fig 6: *The UV-vis and NIR spectra of nitrogen and oxygen*

### III. Principal Component Analysis of Spectra

The effect of the uniform magnetic field on the spectra of nitrogen and oxygen spectra is studied by one of the pattern recognition techniques of principal component analysis (PCA). It is one of the most common multivariable techniques which have been used to analyze the structure of high dimensional data. It reduces the dimension of the data by projecting it onto a space spanned by the vectors called Principal Components (PS). These vectors obtained successively and corresponds to the maximum variance of the remaining data in each time. PCA is widely used in astronomy, medicine, robotics, remote sensing, spectroscopy of astrophysical plasmas and laboratory etc [5, 6].

Now we will give the summary of the mathematical background of PCA and details of the PCA can be found in [9, 10]. Let $\{\Gamma_i\}_{i=1}^{M}$ denote the set of *N*×1 vectors of initial data and $\mu$ be the mean of the data. If $\Phi_i = \Gamma_i - \mu$ for i=1, 2,.., M then the covariance matrix is

$$C = \frac{1}{M}\sum_{i=1}^{M} \Phi_i \Phi_i^t$$

where superscript **t** denotes the transpose. **C** is an *N*×*N* symmetric matrix, Therefore it is diagnosable and has *N* non-negative eigenvalues. Let denote the eigenvectors corresponding to the largest three eigenvalues (from larger to smaller) as |PC1>, |PC2>, and |PC3> .

In this work, PCA is applied to the data obtained for magnetic field of B=0 and B=100 Gauss separately. For each of these cases, 5 different pressure level with 30 spectra is considered. Hence PCA is applied to in total 5x30=150 spectra of size 2825x1 of the case B=0 and B=100 separately. Each obtained principal component is of size 2825x1.

For each spectrum, |PC1>, |PC2> and |PC3> coefficients are obtained by projecting the spectra onto the space spanned by these three orthogonal vectors. Hence each of 150 spectra is represented in a 3 dimensional vector space. In figure.8 |PC1>, |PC2> and |PC3> coordinates of experimental spectra with the presence and absence of magnetic fields are presented and figure 7 shows that the plasma particles becomes coordinated when the magnetic field is present for the oxygen plasma.

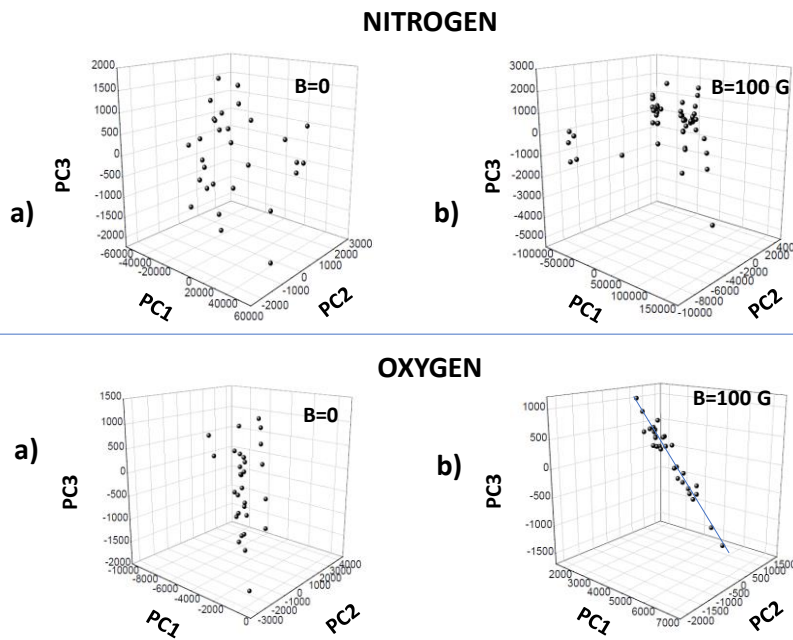

Fig. 7 |PC1>, |PC2> and |PC3> coordinates of spectra in the absence and presence of magnetic fields for nitrogen and oxygen.

In figure 8, |PC1> spectra of nitrogen and oxygen are presented. Spectra show that the intensity of the oxygen is enhanced and the noise data is averaged by the PCA. Oxygen species in the UV-vis and NIR and nitrogen species with transition 337 nm show the linear polarization Stokes profiles by introducing small magnetic fields [13]. Transitions of 337 nm of nitrogen and 845 nm of oxygen are also known to be sensitive to the stimulated emission by lasers in the atmosphere, and they can provide information about pollutants for atmospheric remote sensing [14, 15].

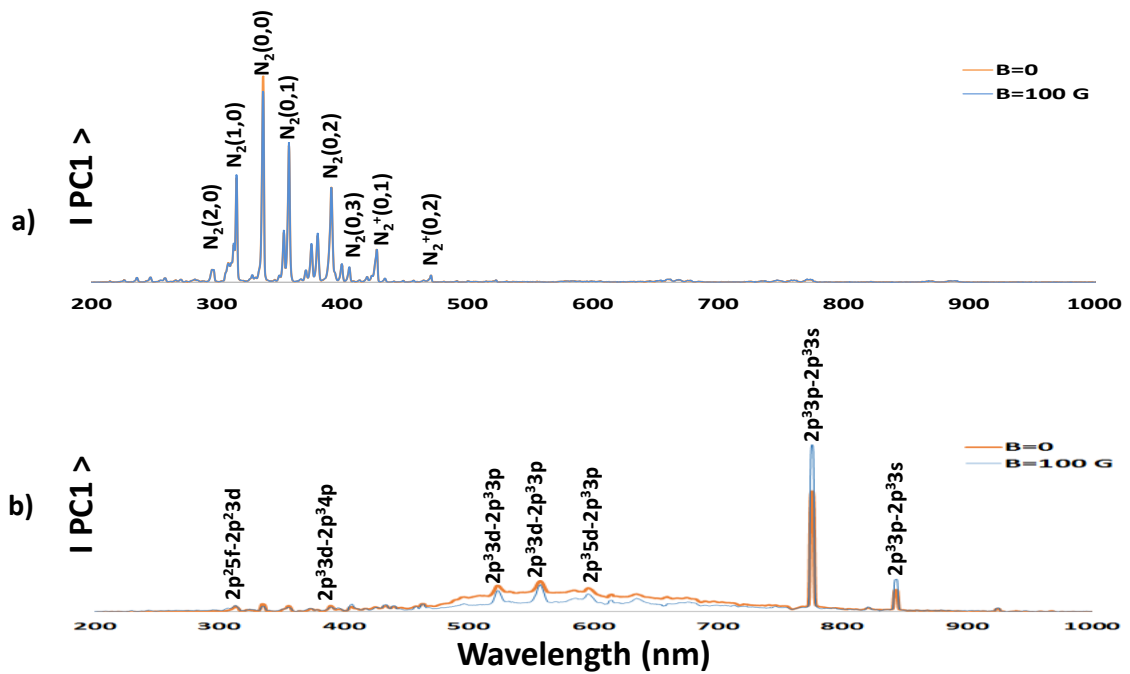

Fig.8 |PC1> spectra of a) nitrogen and b) oxygen plasma in the absence and presence of magnetic field of 100 G

### IV. Non local thermal equilibrium modelling of Nitrogen and Oxygen plasma

The simulation of the experimental spectra of nitrogen and oxygen plasmas is implemented by the SPARTAN code. The SPARTAN is a line-by-line numerical code which calculates the spectral dependent emission and absorption coefficients of a gas which can be either in thermal equilibrium or not. For our modeling, a non-thermal modeling with Doppler broadened line profiles is considered due to nature of glow discharge plasmas. In figure 10, the modelings of nitrogen and oxygen are illustrated. Modelings of nitrogen and oxygen suggest that the plasma electron temperatures are 25000 and 5000 Kelvin, respectively [16].

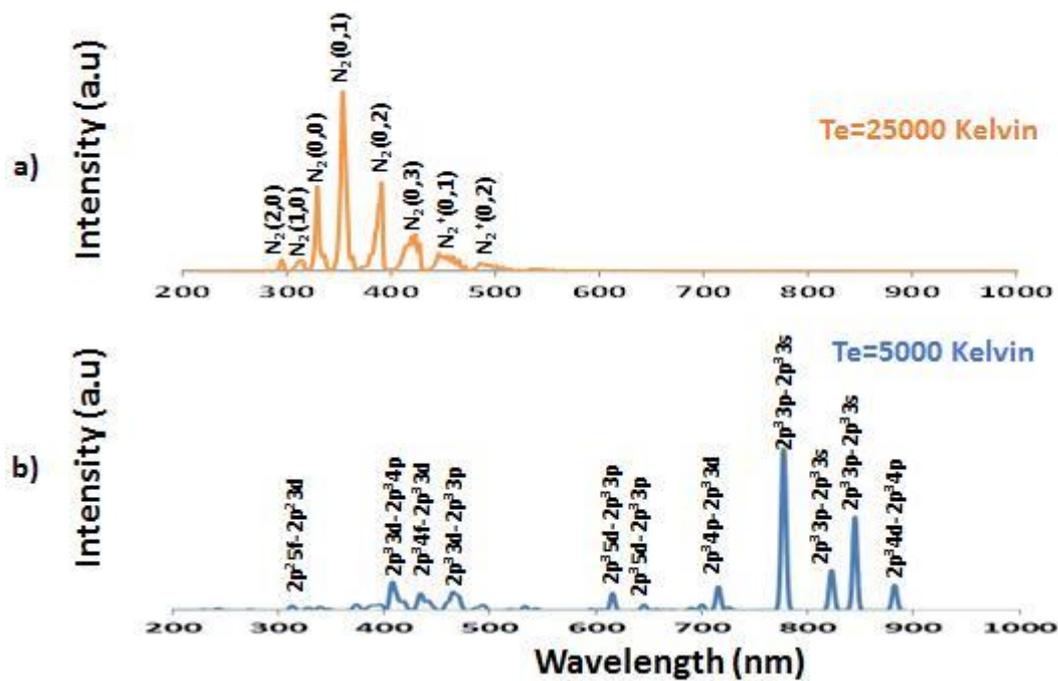
Fig 10 *The modelling of spectra of nitrogen and oxygen*

## V. Discussion

One can basically state the regions of glow discharge plasmas as cathode fall, negative glow, faraday dark space and anode glow. Negative glow region has the highest light intensity and the lowest electric field compared to the other regions. Another characteristic of negative glow region is that it has higher charge density. Besides, Kazantsev et al., 1995 stated that the maximum polarization degree was near the boundary negative glow [17, 18]. Figure 6 shows that oxygen plasma has larger glow of the cathode side compared to the nitrogen, and nitrogen has larger and brighter volume of positive glow. So the larger volume glow of cathode side might be the reason for the linear polarization of the oxygen species in the presence of magnetic fields. Since the electric field component of the radiation is very low in this region, the intensity enhancement might be due to the magnetic field component of the radiation which is parallel to the external magnetic field. Another approach of the enhancement of the intensity of the radiation and strong electric fields are known to be due to the stability of the plasma condensation [19]. According to Gorbachev et al., 1988 plasma condensation is observed usually in coronal plasmas and realized when magnetic force dominates the gravitational and the pressure forces [20]. Fig.7 shows that the linearization of the plasma species might be the signature of the plasma condensation which has been observed in many ion sputtering experiments [21]

**CONCLUSION**

The breakdown voltages of plasma of nitrogen and oxygen were produced by DC current (5 mA, 100-1100 V), while the magnetic field was present (B=100 Gauss), and the gas pressure was changed between (0.1 - 1 Torr). It is observed that the breakdown voltages curve is raised by the presence of magnetic field and magnetic field is more effective on pulse timing of oxygen plasmas. The emissions spectra of nitrogen and oxygen gases are compared for both breakdown voltages with crossed magnetic fields. Each optical emission line is raised by the magnetic field for the oxygen species in UV-Vis and NIR region and only one transition of nitrogen at 337 nm. The present results are useful for data interpretation of optical spectra in plasma diagnostic studies, and PCA is very efficient technique to study the small magnetic field effects and noise reducing on the spectra of discharge plasmas. PCA extracts the effects of magnetic field from the emission spectra in a way of condensed behavior. PCA spectra extract the intensity enhancements of oxygen species and linear polarization of Stokes profiles. Non thermal plasma modeling suggests that plasma electron temperatures of Nitrogen and Oxygen $T_e=25000$ and $T_e=5000$ Kelvin respectively.

## ACKNOWLEDGEMENTS

This research was funded by The Scientific and Technological Research Council of Turkey with the project number of Tubitak-EEAG-113E097

**LIST OF REFERENCES**